\documentclass[prl,aps,twocolumn,showpacs,superscriptaddress,10pt]{revtex4-1}

\usepackage{graphicx}
\usepackage{color}
\usepackage{amsfonts}
\usepackage{cancel}
\usepackage{xcolor}

\usepackage{amsmath, amssymb, amsfonts, stmaryrd}
\usepackage[colorlinks=true,linkcolor=black,citecolor=black,filecolor=black,urlcolor=black]{hyperref}

\newcommand{\hide}[1]{}

\begin{document}

\title{Probing cage relaxation in concentrated protein solutions by XPCS}

%
%

\author{Yuriy Chushkin$^{\dagger}$}
\email{chushkin@esrf.fr}
\thanks{These three authors contributed equally}

\affiliation{ESRF, The European Synchrotron, 71 Avenue des Martyrs, CS40220, 38043 Grenoble Cedex 9, France}

\author{Alessandro Gulotta}
\thanks{These three authors contributed equally}
\affiliation{Division for Physical Chemistry, Lund University, Naturvetarvägen 14, 22100 Lund, Sweden}

\author{Felix Roosen-Runge}
\thanks{These three authors contributed equally}
\affiliation{%
Department of Biomedical Sciences and Biofilms-Research Center for Biointerfaces (BRCB), Faculty of Health and Society, Malm\"o University, Sweden}
\affiliation{Division for Physical Chemistry, Lund University, Naturvetarvägen 14, 22100 Lund, Sweden}

\author{Antara Pal}
\affiliation{Division for Physical Chemistry, Lund University, Naturvetarvägen 14, 22100 Lund, Sweden}
\author{Anna Stradner}
\affiliation{Division for Physical Chemistry, Lund University, Naturvetarvägen 14, 22100 Lund, Sweden}
\affiliation{Lund Institute of advanced Neutron and X-ray Science LINXS, Lund University, Lund, Sweden}
\author{Peter Schurtenberger}
\email[]{peter.schurtenberger@fkem1.lu.se}
\affiliation{Division for Physical Chemistry, Lund University, Naturvetarvägen 14, 22100 Lund, Sweden}
\affiliation{Lund Institute of advanced Neutron and X-ray Science LINXS, Lund University, Lund, Sweden}

\date{\today}

%
%

\begin{abstract}
Diffusion of proteins on length scales of their own diameter in highly concentrated solutions is essential for understanding the cellular machinery of living cells, but its experimental characterization remains a challenge. While X-ray photon correlation spectroscopy (XPCS) is currently the only technique that in principle allows for a measurement of long-time collective diffusion on these length scales for such systems, its application to protein solutions is seriously hampered by radiation damage caused by the highly intense X-ray beams required for such experiments. Here we apply an experimental design and an analysis strategy that allow us to successfully use XPCS experiments in order to measure collective long-time cage relaxation in highly crowded solutions of the eye lens protein $\alpha$-crystallin close to and beyond dynamical arrest. We also address the problem of radiation-induced damage in such experiments. We demonstrate that radiation effects depend both on the total dose as well as the dose rate of the absorbed radiation, and discuss possible processes and mechanism responsible for the observed radiation effects as well as their consequences for future applications of XPCS in biological systems.
\end{abstract}

%
%


\maketitle

%
%

Molecular processes leading to dynamical arrest in biomacromolecular solutions are vital for life, governing protein assembly and condensation as essential formation pathways of biological structure~\cite{Gunton2007}. As examples, both the attractive gel~\cite{Bucciarelli2016} and repulsive glass transition~\cite{Foffi2014} observed in concentrated protein solutions present relevant cases for inter alia optimization of pharmaceutical formulations.

To approach the questions of phase behavior and dynamics of globular protein solutions it is convenient to use experimental and theoretical tools of colloid science~\cite{Stradner2020}. In colloid and glass physics, the mechanistic understanding of dynamical arrest such as during glass formation or gelation has been established by complementing information on multi-scale structure with results on macroscopic viscosity and local diffusion~\cite{Hunter_2012}. In particular, the microscopic information on caging processes is key to reaching a mechanistic understanding how local stresses in the sample are distributed and released~\cite{Hunter_2012}.

Since the nonspherical protein conformations with anisotropic molecular interactions render naive colloidal interpretations problematic~\cite{Stradner2020}, a comprehensive experimental characterization of protein solution is essential for a quantitative, mechanistic understanding, theory testing and development. While multi-scale structural information and macroscopic dynamics are accessible using imaging and scattering techniques~\cite{McManus2019}, diffusion on the length scale of nearest neighbors is only available on short time scales up to a few hundred nanoseconds by neutron spin-echo spectroscopy, corresponding to typical rattling-in-the-cage motion~\cite{Roosen_2011,Bucciarelli2016,Grimaldo_2019a,Grimaldo_2019}.

Information on longer micro- to millisecond time scales addressing out-of-cage diffusion is so far missing. Given the length scale of roughly 10 nanometers for proteins, the only experimental technique promising to probe the collective density relaxations is X-ray photon correlation spectroscopy (XPCS)~\cite{Grubel2008,Vodnala2018}. 
However, so far applications to biological matter have been hampered by the X-ray radiation damage~\cite{Vodnala2018,Moller2019}. The conventional approach to cope with the radiation damage has been to limit the deposited dose below a critical value~\cite{Vodnala2018,Lurio2021}, before structural changes become measurable. Low dose XPCS requires distribution of the radiation over a large sample volume to optimize the scattered signal~\cite{Moller2019}. For example,  using a 100$\times$100 $\mu$m large beam to distribute the dose, collective dynamics at long length scales could be addressed e.g.~in concentrated antibody formulations~\cite{Girelli2021}. Alternatively, one can translate a sample and control the cumulative dose by the translation speed~\cite{Vodnala2018,Lurio2021}. By this approach, proof-of-principle results on the cage relaxation in concentrated solutions of the protein $\alpha$-crystallin were reported~\cite{Vodnala2018}, but relaxation times were limited to the translation times, and thus do not extend to the very long time scales relevant in concentrated samples close to dynamic arrest. Moreover, recent work~\cite{Lurio2021} reported a five times lower critical dose for the same protein than in the earlier study~\cite{Vodnala2018}. It is clear that we lack sufficient fundamental understanding of the role of radiation effects on XPCS experiments with highly concentrated protein solutions. In this letter we therefore address the problem of radiation-induced damage for such experiments, and show that collective long-time cage relaxation in highly crowded protein solutions measured with XPCS can be affected by the dose rate even below critical dose values. We characterize the beam induced dynamics and demonstrate that by using a specific experimental design and analysis strategy we can indeed correctly estimate the intrinsic solution dynamics in agreement with existing rheology data. Furthermore, we make an attempt to elucidate the role of the underlying processes and mechanism responsible for the observed radiation effects. 

In this study, we benefit from a $\sim$100 fold increase in the coherent fraction of the recently upgraded extremely brilliant source (EBS) at the ESRF\cite{Raimondi2016}. The SAXS and XPCS measurements were performed at the ID10 beamline. We used 9.5 keV radiation selected by a Si(111) channel-cut monochromator and beam sizes of 20x20 or 30x30 $\mu$m$^2$ with high degree of spatial coherence. The maximum intensity was 3.74$\times 10^{11}$ for the focused and 2.45$\times 10^{10}$ ph/s for the unfocused beam. Si attenuators of different thicknesses were inserted to reduce the photon flux and hence the dose rate and dose of the  absorbed radiation. The deposited dose was calculated following~\cite{Kuwamoto2004}. Series of scattering patterns from a sample were recorded by the single-photon-counting Eiger 500K pixel array detector \cite{Zinn2018} placed 6.9 m downstream from a sample and shifted to access the static structure peak.
 This setup with a very high coherent fraction provides ideal conditions to first systematically explore effects of dose rate and cumulative dose, and second test whether physical information on collective solution dynamics can be reliably extracted.

We used concentrated solutions of the protein $\alpha$-crystallin from the bovine eye lens as a  well-established model system. Importantly, $\alpha$-crystallin has been studied in detail using static and dynamic light scattering, small-angle X-ray scattering, neutron spin echo and macro- and micro-rheology \cite{Foffi2014,Bucciarelli2016,Garting2018}, all of which support a coherent picture for the solution structure, collective diffusion and the glass transition based on a model of polydisperse hard spheres. The use of XPCS on this system provides us thus with the ideal experiment to on the one hand complete the experimental physical characterization and also obtain the missing information on long-time cage relaxation, and on the other hand test the potential of XPCS for crowded protein solutions.

We prepared a series of concentrations following a well-established protocol \cite{Bucciarelli2016} in a concentration range between 300 mg/mL and 360 mg/mL, corresponding to a protein volume fraction $\phi$ between 0.51 and 0.61 \cite{Foffi2014}. As a solvent we use a 52.4 mM phosphate buffer (pH 7.1) with 20 mM DTT (dithiothreitol) and 1 mM EDTA (ethylenediaminetetraacetic acid) to reduce effects of oxidation stress and radiation damage. The samples were sealed in quartz capillaries with 1.5 mm diameter, and measured at room temperature 23$^{\circ}$C.
We used $\sim$10 to $\sim$1000 fresh spots per sample with 60 $\mu$m spacing, and collected series of 500 to 20000 frames for each spot to achieve good signal-to-noise ratios. For each sample several data sets with different beam intensity and acquisition time were acquired.

	\begin{figure}
		\includegraphics[width=\linewidth]{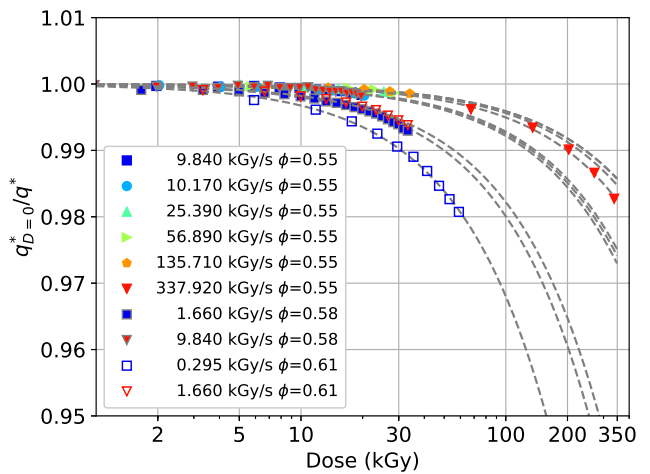}
    				\caption{ 
    					Inverse of the relative position $q^*_{D=0}/q^*$ of the peak in the structure factor $S(q)$, where $q^*_{D=0}$ is the extrapolated zero-dose limit,  as a function of deposited dose and dose rate for $\alpha$-cystallin solutions at different volume fractions $\phi$. Up to 10 kGy cumulative dose, the relative values barely change, independent of dose rate and volume fraction. For higher doses, clear signatures of radiation-induced damage are observed.
				 The guide to the eye (gray dashed) is an exponential decay.}
		\label{fig:structure}
	\end{figure}

All samples show a clear correlation peak in the small-angle scattering curve (SI Fig. S1 (a)). Fitting this by the product of the low concentration form factor of $\alpha$-crystallin and a structure factor for a polydisperse hard-sphere model~\cite{Vrij1979}, we extract the peak height $S(q^*)$ and the peak position $q^*$ to monitor beam-induced structural changes in a sample. Fig.~\ref{fig:structure} shows the inverse of the extracted parameter $q^*$ normalized with the zero-dose limit $q^*_{D=0}$ versus the deposited dose. In the calculation of $q^*_{D=0}$ we assumed an exponential decay (dashed lines). With increasing dose the peak height decreases (SI Fig. S1 (b)) and the peak position shifts to larger values of the scattering vector $q$. Importantly, the rate of change is not constant but depends on the dose rate and volume fraction. This behavior explains the earlier conflicting findings on critical doses for $\alpha$-crystallin \cite{Vodnala2018,Lurio2021}.

	\begin{figure*}
		\includegraphics[width=\linewidth]{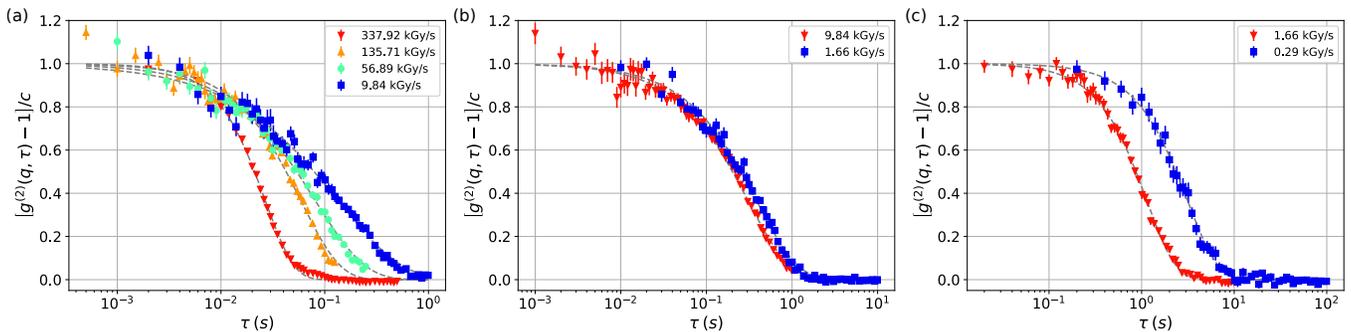}
				\caption{Measured normalized intensity correlation functions from $\alpha$-crystallin solutions at different dose rates and volume fractions (a) $\phi$=0.55, (b) $\phi$=0.58 and (c) $\phi$=0.61. Dashed lines are fits with Eq.~\eqref{eq:kww}. We remark that the data quality outlines effects of dose rate already without numerical analysis.}
		\label{fig:gqt}
	\end{figure*}
	
\begin{figure*}
		\includegraphics[width=\linewidth]{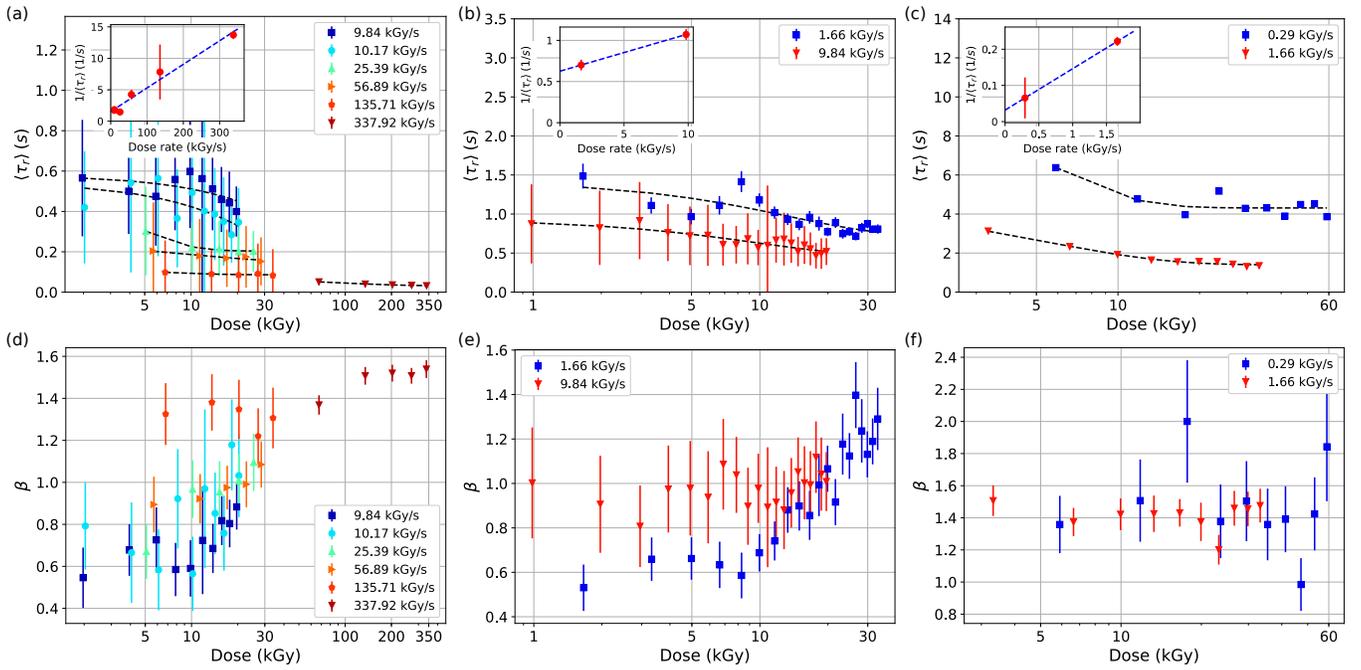}
		
		\caption{Observed dynamical characteristics of $\alpha$-crystallin solutions as a function of the cumulative dose for different dose rates and three volume fractions: $\phi$=0.55 (left), $\phi$=0.58 (center), $\phi$=0.61 (right). Top row: the average relaxation time $\left \langle \tau_{r} \right \rangle$ is decreased by both dose rate and dose. Inset: the relaxation rate as a function of the dose rate. Bottom row: the exponent parameter $\beta$ evidences a transition from stretched exponential decay in the weakly affected fluid state to compressed decay signatures for the strongly affected state.}
		
		\label{fig:dynamics}
	\end{figure*}

Indeed, the measured intensity correlation functions $g^{(2)}(q^*,t)=\frac{\left< I(q^*,0)I(q^*,t)\right>}{\left< I(t) \right>^2}$ shown in Fig.~\ref{fig:gqt} for different volume fractions are sensitive to the dose rates. Without further analysis, a significant effect of dose rate on the decay time and shape is observed, even if the absolute dose is significantly below 10 kGy (Fig.~\ref{fig:gqt} (a)).
To quantify this observation, we fitted the correlation functions with the Kohlrausch-Williams-Watts (KWW) expression~\cite{Williams1970}
\begin{equation}g^{(2)}(q,\tau)=1+c\cdot\exp(-2\cdot(\tau/\tau_{r})^{\beta})\ .
\label{eq:kww}
\end{equation}
Here, $c$ is the contrast of the correlation function, and $\tau_{r}$ is the relaxation time. The contrast $c$ was higher than 5\% for a 30$\times$30 $\mu m$ beam. The dose dependent behavior is analyzed using time resolved $g^{(2)}$ extracted from the two-time correlation function~\cite{Skouripanet2006} shown in SI Fig. S3. The relaxation exponent $\beta$ reflects the signature of the dynamical process. A stretched exponential decay is modeled by $\beta<1$, a compressed function has $\beta>1$ and for a simple exponential decay $\beta=1$. 
 Fig.~\ref{fig:dynamics} summarizes $\beta$ and the average relaxation time 
    \begin{equation}
    \left\langle \tau_{r} \right\rangle =\frac{\tau_{r}}{\beta}\Gamma\left(\frac{1}{\beta}\right)\ .
    \label{eq:1}
    \end{equation} 
 for different dose rates and volume fractions.
\newcommand{\tauave}{$\langle \tau_{r}\rangle$ }
Both \tauave and $\beta$ are affected by cumulative dose and dose rate (Fig.~\ref{fig:dynamics}). While at low doses, below the radiation damage threshold, we observe slow dynamics and a stretched exponential, large dose rates lead to faster dynamics, and a transition from stretched to compressed signatures. We remark that the dynamical signatures depend even stronger on dose rate than the structural features in Fig.~\ref{fig:structure}. This becomes evident when comparing the estimated decorrelation related to the shift of $q^{*}$ (SI Fig. S2) with the measured decay of $g^{(2)}(q,\tau)$, which shows that the temporal evolution of the structural features is significantly slower than the beam-induced dynamics observed in XPCS.

Interestingly, the sensitivity of the samples to radiation-induced effects seems related to the physical state as outlined by the examples shown in Fig.~\ref{fig:gqt}. The sample at a volume fraction of $\phi = 0.55$ is close to the glass transition but still fluid, and should thus show a stretched relaxation with $\beta<1$ 
\cite{Bartsch1992, Sciortino2005}. Indeed, both \tauave and $\beta$ show consistent and stable values below 10 kGy for the smallest two dose rates around 10 kGy/s. Larger dose rates induce faster relaxations of different nature with $\beta>1$ (Fig.~\ref{fig:dynamics} (a,d)).

The sample at the intermediate volume fraction of 0.58 is at the arrest transition, which should again result in a correlation function that follows a stretched exponential decay. While the measurement at the lowest dose rate returns a reasonable $\beta<1$, a dose rate around 10 kGy/s already induces an inconsistent $\beta\approx 1$, implying a higher sensitivity to dose rate than for the slightly less concentrated sample.

The highest volume fraction of 0.61 corresponds to an arrested or non-ergodic sample, which would in principle imply a constant plateau for the long-time $\alpha$ relaxation observed by XPCS. However, for our sample structural relaxation appears to be induced already for much smaller dose rates around 1 kGy/s, with $\beta \approx 1.5$ and a \tauave that is of similar magnitude as the values obtained at the arrest transition. In addition, \tauave does not show a constant value below 10 kGy, suggesting that the dynamical signatures observed are related to radiation-induced relaxation processes. For typical colloidal hard or soft sphere glasses one generally indeed finds a final very slow decay in the correlation functions due to ageing processes that result in a compressed exponential decay with $\beta \approx 1.5$, and which are believed to result from the relaxation of internal stresses or strain through non-diffusive processes \cite{Cipelletti2003}. 

\begin{figure}
		\includegraphics[trim=8 0 0 0,clip,width=\linewidth]{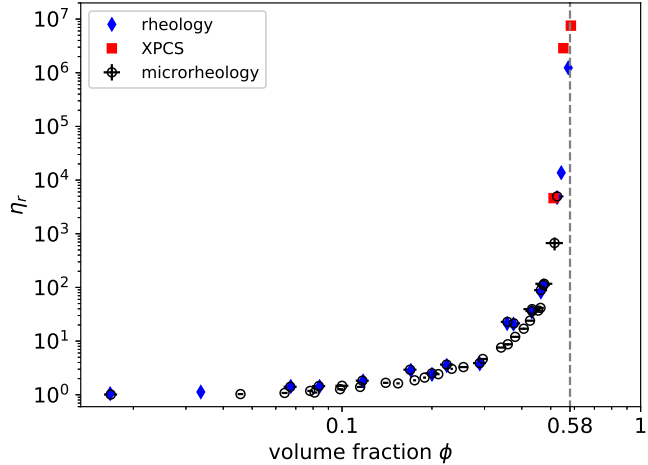}
				\caption{Relative zero-shear viscosity $\eta_{r}$ of $\alpha$-crystallin solutions as a function of volume fraction $\phi$ obtained by different techniques.
				The XPCS data extracted from the intrinsic dynamics in this study (red squares) are in perfect agreement with earlier measurements using rheology (data from \cite{Foffi2014}) and microrheology (data from \cite{Garting2018}).}
		\label{fig:visc}
	\end{figure}

Before attempting to comment on the origin of the radiation-induced temporal evolution observed with the different samples, we investigate whether we can correct for the effect of dose rate and extract the correct intrinsic long-time collective dynamics. Based on the experimental fact that the beam-induced dynamics and the intrinsic dynamics are two independent and parallel relaxation processes~\cite{Ruta2017,Pintori2019} we determine the average relaxation rate $1/\left\langle \tau_{r} \right\rangle$ (at zero dose) versus dose rate (Insets in Fig.~\ref{fig:dynamics}) as proposed in~\cite{Chushkin2020}. The observed linear relationship allows extrapolation to the limit of zero dose rate for our samples, which approaches the true intrinsic dynamics of the system~\cite{Chushkin2020}. To verify the validity of this approach we need to compare the estimated values with complementary measurements using other techniques. Typically, the long-time $\alpha$ relaxation scales with the macroscopic viscosity of the solution. Fig.~\ref{fig:visc} displays viscosities from earlier rheology \cite{Foffi2014} and microrheology \cite{Garting2018} measurements along with the estimated relaxation rates normalized by the dilute limit $\Gamma_0=D_0q^2$ with the diffusion coefficient $D_0=2.2\times10^{-11}$ m$^{2}$/s. We obtain good agreement, which validates the applicability of XPCS for obtaining physical information of the undisturbed sample, if systematic corrections for dose-rate effects are performed.

What remains to be discussed is the physical origin of the intriguingly rich dynamical phenomenology observed in our XPCS experiments: while a stretched-exponential signature is observed in high-concentration, still fluid protein solutions at low dose rates, the increase of both dose rate and the protein volume fraction results in compressed-exponential signatures. We remark that a transition from stretched to compressed correlation functions with a decrease of the temperature or an increase in volume fraction was reported in metallic~\cite{Ruta2012} and colloidal~\cite{Kwasniewski2014} glasses, respectively. However, the dependence of dose rate does not change the relaxation signature in these systems, and thus seems to be a new observation in these proteins solutions, likely also relevant for other radiation-sensitive biological systems.

As a likely explanation, we consider the formation of covalent links between proteins caused by radiation-induced free radicals in the solvent~\cite{Weik2000,Jeffries2015}.
Upon illumination with X-rays, highly reactive hydroxyl (HO·) radicals are formed by the decomposition of water as the dominant process~\cite{Gebicki2021}. In concentrated protein solutions, HO· almost instantaneously diffuses to and reacts with the protein surface into C-centered protein radicals. When coming in close contact, these protein radicals react, and form permanent covalent bonds between protein molecules \cite{Hawkins2001,Jeffries2015}. 
Typically, protein cluster formation results in a slowing down of diffusion, and a concomitant appearance of a stretched exponential decay of the correlation
function. However, we stress that our experiments do not study the equilibrium dynamics of protein clusters, but the non-equilibrium situation of protein solutions during cluster formation.
The non-equilibrium nature of the underlying radical reaction induces changes in the free energy landscape during bond creation. In particular, the close binding of two proteins leads to the opening of a local void in the nearest neighbor cages of the crowded protein solutions, into which the surrounding proteins relax with a locally directed motion. Thus, the driven process of radical bond formation is expected to induce an apparent quasi-ballistic collective motion explaining the compressed signature in the collective relaxation.

Importantly, the relevance of these processes on the experimental time scales can be supported by a quantitative estimation: the formation rate of radicals per protein molecule can be expressed as
\begin{align}
k_r = G j \frac{\rho_{H_2O} V_{beam}}{N_p} \approx \frac{0.68}{s}\frac{j}{kGy/s}
\end{align}
where $j$ is the equivalent dose rate, $G = 2.8 \cdot 10^{-7}$ mol J$^{-1}$ expresses the number of hydroxyl radicals per absorbed radiation dose \cite{Gebicki2021}, $\rho_{H_2O}\approx 1$g/ml is the water density, and $V_{\rm beam} = 30\, \mu\text{m} \cdot 30\, \mu\text{m}\cdot 1.5$\,mm is the illuminated beam volume. The number of proteins $N_p = V_{\rm beam}c_p/M_w \approx 5.6\cdot 10^{-13}$ mol has been evaluated for realistic values for the  protein mass concentration $c_p \approx 330$\,mg/ml, and a molecular weight of $M_w\approx 800$ kDa. 
Thus, for a dose rate of 10 kGy/s, the proteins obtain on average one radical every $\tau_r =1/k_r \approx 150$\, ms.

The experimental condition of a protein in a dense cage of other proteins leads to a situation in which rotational tumbling and translational collision times are on microsecond time scales, while out-of-cage diffusion is orders of magnitudes slower. The first radical bonds thus form with a rate $k_b\approx 10^5$/s as soon as two neighboring proteins receive one radical each. The basic rate equations for the average number of free radicals per protein $n_r$ as well as the number of radical bonds $n_b$ read in a mean-field approximation
\begin{align}
\frac{dn_r}{dt} &= k_r - 2 \frac{dn_b}{dt}\\
\frac{dn_b}{dt} &= k_b n_r^2
\end{align}
These equations are solved by $n_r(t) = \sqrt{k_r/(2k_b)}\tanh(\sqrt{2 k_r k_b}t)$ and $n_b(t) = 0.5 k_r t - 0.5 n_r(t)$, which implies an average time for one bond formation per protein of $\tau_b = 2/k_r \approx 300$ ms.

It is important to stress that the formation of subsequent bonds will be slowed down considerably, as the rotation and translation of proteins needed to find neighboring radicals becomes significantly obstructed, once clusters are formed. Our simple consideration thus provides a rationale why non-equilibrium driven motions induced by X-ray radiation should be expected in these dense proteins on time scales of 100 ms to seconds. 

For liquid samples close to dynamical arrest, this proposed mechanism thus explains why the long time relaxation of $g^{(2)}(q,\tau)$ changes from a stretched to a compressed exponential behavior, while for the already arrested non-ergodic samples all correlation functions measured are compressed with a characteristic relaxation time that is accelerated with increased absorbed radiation. It is also consistent with our observations that the position $q^*$ of the nearest neighbor peak in $S(q)$ increases and the corresponding peak amplitude $S(q^*)$ decreases with increasing time or total dose, as these quantities reflect the increasing number of defects created in the cage structure caused by radical bond formation between proteins and the concomitant decrease of the average nearest neighbor distance.

We remark that this picture might be different for lower volume fractions, where radiation-induced clustering has been speculated about based on radiation-induced slowing down~\cite{Vodnala2016, Reiser2022}. Under these conditions we expect that radiation-induced aggregation would exhibit the typical features of reaction-limited cluster-cluster aggregation \cite{Lin1989} and strongly interfere with our ability to measure collective dynamics on longer time scales with XPCS. In contrast, at very high concentrations close to dynamical arrest, the proposed mechanism is instead related to percolation processes \cite{Stauffer1994}. Here the large difference between short-time diffusion related to rattling in the nearest neighbor cage and long-time diffusion caused by cage relaxation only allows for bond formation between proteins on nearest neighbor positions, and diffusion of clusters over longer distances is virtually prohibited by the cages.

We furthermore remark that the beam-induced dynamics is of non-thermal nature, as from finite difference calculations for the highest dose and dose rate we obtain a maximum heating of only $\sim$0.44$^{\circ}$C (SI Fig. S4 (b)). As such the effect is similar to the acceleration of the structural relaxation with dose rate observed in 
silica glass~\cite{Ruta2017} and in other oxide glasses~\cite{Pintori2019,Holzweber2019}, and explained by breaking and reformation of atomic bonds~\cite{Ruta2017,Pintori2019,Chushkin2020} as a nonthermal effect~\cite{Medvedev2015}.

In conclusion, using the highly coherent X-ray beam of the EBS source we could collect high-quality XPCS data on concentrated $\alpha$-crystallin solutions. By applying a systematic approach - varying dose and dose rate - we could reveal information on the physical mechanism behind dynamical arrest on nearest-neighbor length scale that agrees with and completes the previous studies. Moreover, we obtain indirect information on the transient-bond mechanism behind structural changes and dynamical acceleration induced by X-ray radiation. The results show that in addition to the critical dose the critical dose rate is the relevant parameter to control in the XPCS measurements. These findings imply that access to the true microscopic dynamics of biological samples is possible, if experiments stay both below the critical deposited dose and the critical dose rate. Importantly, the critical dose rate seems to be related to the ratio of the critical dose to the relaxation time of the system. Therefore, an experimental characterization of short-time cage diffusion may even be possible with XPCS when using sufficiently intense X-ray beams where the absorbed dose would be above the typical estimate of the critical dose of 10 kGy, but where bond formation would be sufficiently delayed due to the significant difference between short time diffusion and bond formation. Given the enormous improvements of coherent beam characteristics at modern synchrotron sources, XPCS in this spirit is highly promising for future in-depth characterization of local dynamics in biological and soft materials. 

This research was financially supported by the Swedish National Research Council (Vetenskapsrådet)
under Grant No. 2016-03301, the R{\"{o}}ntgen-Ångstr{\"{o}}m Cluster Grant No. 2019-06075, and the Crafoord Foundation (Lund, Sweden). We acknowledge support by LINXS - Lund Institute of Advanced Neutron and X-ray Science. Md. Arif Kamal, T. Narayanan and T. Zinn are acknowledged for fruitful discussions and help. The authors would like to thank F. Zontone for technical assistance during the experiment. We acknowledge the European Synchrotron Radiation Facility for provision of synchrotron radiation facilities for experiment number SC5047.

\bibliography{protein_dynamics.bib}

\end{document}